\pdfoutput=1
\RequirePackage{ifpdf}
\ifpdf 
\documentclass[pdftex]{sigma}
\else
\documentclass{sigma}
\fi

\def\PP{\mathbb{P}}

\def\KK{\mathbb{K}}
\def\CC{\mathbb{C}}

\begin{document}

\allowdisplaybreaks

\renewcommand{\PaperNumber}{041}

\FirstPageHeading

\ShortArticleName{$N$-Qubit and $2^{N - 1}$-Qubit Pauli Groups via Binary ${\rm LGr}(N,2N)$}

\ArticleName{A Notable Relation between $\boldsymbol{N}$-Qubit and\\ $\boldsymbol{2^{N - 1}}$-Qubit Pauli Groups via Binary $\boldsymbol{{\rm LGr}(N,2N)}$}

\Author{Fr\'ed\'eric HOLWECK~$^\dag$, Metod SANIGA~$^\ddag$ and P\'eter L\'EVAY~$^\S$}

\AuthorNameForHeading{F.~Holweck, M.~Saniga and P.~L\'evay}

\Address{$^\dag$~Laboratoire IRTES/M3M, Universit\'e de Technologie de Belfort-Montb\'eliard,\\
\hphantom{$^\dag$}~F-90010 Belfort, France}
\EmailD{\href{mailto:frederic.holweck@utbm.fr}{frederic.holweck@utbm.fr}}

\Address{$^\ddag$~Astronomical Institute, Slovak Academy of Sciences,\\
\hphantom{$^\ddag$}~SK-05960 Tatransk\' a Lomnica, Slovak Republic}
\EmailD{\href{mailto:msaniga@astro.sk}{msaniga@astro.sk}}

\Address{$^\S$~Department of Theoretical Physics, Institute of Physics,
Budapest University\\
\hphantom{$^\S$}~of Technology and Economics, Budafoki \'ut.~8, H-1521, Budapest, Hungary}
\EmailD{\href{mailto:levay@neumann.phy.bme.hu}{levay@neumann.phy.bme.hu}}

\ArticleDates{Received November 14, 2013, in f\/inal form April 02, 2014; Published online April 08, 2014}

\Abstract{Employing the fact that the geometry of the $N$-qubit ($N \geq 2$) Pauli group is embodied in the structure of the symplectic polar space $\mathcal{W}(2N-1,2)$ and using properties of the Lagrangian Grassmannian ${\rm LGr}(N,2N)$ def\/ined  over the smallest Galois f\/ield, it is demonstrated that there exists a bijection between the set of maximum sets of mutually commuting elements
of the $N$-qubit Pauli group and a certain subset of  elements of the $2^{N-1}$-qubit Pauli group. In order to reveal f\/iner traits of this correspondence, the cases $N=3$ (also addressed recently by L\'evay, Planat and Saniga [\textit{J.~High Energy Phys.} \textbf{2013} (2013), no.~9, 037, 35~pages]) and $N=4$ are discussed in detail.  As an apt application  of our f\/indings, we use the stratif\/ication of the ambient projective space ${\rm PG}(2^N-1,2)$ of the $2^{N-1}$-qubit Pauli group in terms of $G$-orbits, where $G \equiv {\rm SL}(2,2)\times {\rm SL}(2,2)\times\cdots\times {\rm SL}(2,2)\rtimes S_N$, to decompose $\underline{\pi}({\rm LGr}(N,2N))$ into non-equivalent orbits. This leads to a partition of ${\rm LGr}(N,2N)$ into distinguished classes that can be labeled by elements of the above-mentioned Pauli groups.}


\Classification{05B25; 51E20; 81P99}

\section{Introduction}
Generalized Pauli groups (also known as Weyl--Heisenberg groups) associated with f\/inite-di\-men\-sio\-nal Hilbert spaces play an important role in quantum information  theory, in particular in quantum tomography, dense coding, teleportation, error correction/cryptography, and the black-hole-qubit correspondence. A~special class of these groups are the so-called $N$-qubit Pauli groups, $N$ being a positive integer, whose elements
are simply $N$-fold tensor products of the famous Pauli matrices and the two-by-two unit matrix. A~remarkable property of these particular groups is that their structure can be completely recast in
the language of symplectic polar spaces of rank~$N$ and order~2, $\mathcal{W}(2N -1,2)$ (see, for example, \cite{hos,pla, ps,slp,sp,spp,th} and references therein).
The elements of the group (discarding the identity) answer to the points of $\mathcal{W}(2N - 1, 2)$, a maximum set
of pairwise commuting elements has its representative in a maximal subspace (also called a~generator) ${\rm PG}(N-1,2)$,  the projective space of dimension~$N-1$ over the Galois f\/ield  of order~$2$,  of $\mathcal{W}(2N - 1, 2)$ and, f\/inally, commuting translates into collinear (or, perpendicular). In the case of the {\it real} $N$-qubit Pauli group, the structure of the corresponding $\mathcal{W}(2N -1,2)$ can be ref\/ined in terms of the orthogonal polar space $\mathcal{Q}^{+}(2N - 1, 2)$, that is, a hyperbolic quadric of the ambient projective space ${\rm PG}(2N - 1, 2)$, which is the locus accommodating all symmetric elements of the group~\cite{hos}. Given this f\/inite-geometrical picture of (real) $N$-qubit Pauli groups, one can invoke properties of the Lagrangian Grassmannian ${\rm LGr}(N,2N)$ def\/ined over the Galois f\/ield of two
elements,
${\rm GF}(2)$, to establish a very interesting bijection between the generators of $\mathcal{W}(2N-1,2)$ and points lying on
a sub-conf\/iguration of $\mathcal{W}(2^N-1,2)$ def\/ined by a set of quadratic equations.
This furnishes an intriguing mapping of maximum sets of mutually commuting $N$-qubit observables into
observables of $2^{N-1}$-qubits.
For $N=3$, all essential technicalities of this relation have recently been worked out in detail in \cite{PMM}. In this paper, we shall
f\/irst give a short rigorous proof that this bijection holds for any $N$. Then, after a brief addressing of a~rather trivial $N=2$ case, we shall again discuss in detail the $N$=3 case using,
however, a more ``projective-slanted'' view to be compared with an ``af\/f\/ine'' approach of the latter reference, as well as the $N=4$ case
to see some novelties and get a feeling of the kind of problems one can envisage/encounter when addressing higher rank cases.

{Our main motivation for having a detailed look at the above-outlined `Lagrangian Grassmannian' relationship between dif\/ferent
multi-qubit Pauli groups stems from an important role of the maximum sets of mutually commuting $N$-qubit observables in the quantum
information theory. On the one hand, such sets are vital for simple demonstrations of} {\it quantum contextuality}.  Every such set
can be regarded as a context and various `magic' collections of such contexts are intimately linked with sub-geometries of the
associated symplectic polar space $\mathcal{W}(2N -1,2)$. The simplest such conf\/iguration can already be found in the $N=2$ case,
being known as a Mermin magic square~\cite{mer}.  It represents a set of nine observables placed at the vertices of a $3 \times 3$
grid and forming six maximum sets of pairwise commuting elements that lie along three horizontal and three vertical lines,
each observable thus pertaining to two such sets. The observables are selected in such a way that the product of their triples
in f\/ive of the six sets is~$+I$, whilst in the remaining set it is $-I$, $I$ being the identity matrix.
Geometrically, each Mermin square is isomorphic to the smallest slim generalized quadrangle, ${\rm GQ}(2,1)$, or
to a hyperbolic quadric~$\mathcal{Q}^{+}(3,2)$. A~number of other magic conf\/igurations, exhibited by higher-order
Pauli groups and featuring a varying degree of complexity, can be found in Waegell's preprint~\cite{wae}.  On the other hand,
existence of these sets is intricately related to the existence of  {\it mutually unbiased bases}
(MUBs) of the associated Hilbert space. In particular, $\mathcal{W}(2N -1,2)$ possesses spreads~\cite{thas},  that is sets
of generators of $\mathcal{W}(2N -1,2)$ partitioning its point-set, whose cardinality is equal to the maximum number
of MUBs, $d+1$, in the associated $d=2^N$-dimensional Hilbert space. Thus, for example, spreads of~$\mathcal{W}(3,2)$
feature f\/ive elements each, and the associated 4-dimensional Hilbert space is indeed found to be endowed with
sets of $4+1=5$ MUBs~\cite{ps}.

The paper is organized as follow. In Section~\ref{symp}, we recall the def\/inition of the symplectic polar space $\mathcal{W}(2N-1,2)$ and how this space
encodes the geometry of the $N$-Pauli group. In Section~\ref{bijection}, we prove our main result by establishing the
existence of a projection which maps bijectively the aggregate of maximum sets of mutually commuting $N$-qubit
observables into a~distinguished subset of $2^{N-1}$-qubit observables. Then, in Section~\ref{example}, we illustrate our construction for  $N = 2, 3$, and~4 by explicitly computing the equations def\/ining the image of
the projection in ${\rm PG}(2^N-1,2)$. In Section~\ref{partition}, one shows how our f\/indings can be used to partition the set of generators of $\mathcal{W}(2N-1,2)$.
Finally, in Section~\ref{variety_minor} we point out a relation between our construction and similar ones done over the f\/ield of complex numbers.

{\bf Notation.}
In what follows, we will denote by $\KK$ the Galois f\/ield ${\rm GF}(2)$ and, if $V$ is a~$\KK$-vector space,  we will use the symbol $\PP(V)$ to represent the corresponding projective space over
$\KK$; thus, $\PP(\KK^{N})$ will be an alternative expression for ${\rm PG}(N-1,2)$, the projective space of dimension $N-1$
over~${\rm GF}(2)$.
Given a nonzero vector $v\in V$, we will denote by $[v]\in \PP(V)$ the corresponding point
in the associated projective space. On the other hand, for any $X\subset \PP(V)$, we def\/ine the cone over~$X$,
$\widehat{X}\subset V$, to be the pre-image of $X$ in $V$, i.e.\ the set of all vectors $x\in V$ such that $[x]\in X$. A~tensorial basis of $(\KK^2)^{\otimes n} \equiv \KK^2\otimes\dots\otimes \KK^2$ ($n$ factors)
will be denoted by $x_1^{i_1}\otimes x_2^{i_2}\otimes \dots \otimes x_n^{i_n}$, where $i_j\in \{0,1\}$; obviously, $\{x_i^0,x_i^1\}$ is a basis of $(\KK^2)_i$.

Let $A=(a_{ij})$ be an $n\times n$ matrix with coef\/f\/icients in $\KK$ and let $I=\{i_1,\dots,i_k\}$ and $J=\{j_1,\dots,j_k\}$ be subsets of $\{1,\dots,n\}$. The symbol $\Delta_{I,J}$ will stand for the corresponding $k\times k$ minor of $A$, i.e.\ $\Delta_{I,J}(A)=\det ((a_{i,j})_{i\in I, j\in J})$;
when $I=J$,
$\Delta_{I,I}(A)$ will be called a~principal minor of $A$ and simply referred to as $\Delta_I(A)$.

In Section~\ref{example}, computations will be handled using
Maple and Macaulay2
to get the equations of the ideal of the Lagrangian Grassmannian ${\rm LGr}(N,2N)$ for $N=3$
and $N=4$. The sources of the codes are available at \url{http://www.emis.de/journals/SIGMA/2014/041/codes.zip}
which contains two f\/iles: one is a Maple f\/ile to compute all equations def\/ining the ideal of
${\rm LGr}(4,8)$, the other is a Macaulay2 script to compute the ideal of the projection of ${\rm LGr}(4,8)$
by elimination theory based on the equations stemming from the previous code.

\section[The  symplectic polar space $\mathcal{W}(2N-1,2)$ and the associated $N$-qubit Pauli group]{The  symplectic polar space $\boldsymbol{\mathcal{W}(2N-1,2)}$\\ and the associated $\boldsymbol{N}$-qubit Pauli group}\label{symp}

 A~(f\/inite-dimensional) classical polar space describes the geometry of a $d$-dimensional vector space
over the Galois f\/ield ${\rm GF}(q)$, $V(d, q)$, carrying a non-degenerate ref\/lexive
sesquilinear form~$\sigma$ (see, e.g.,~\cite{cam}). {The polar space is called symplectic,
and usually denoted as $\mathcal{W}(d - 1,q)$,  if this form is bilinear
and alternating, i.e., if $\sigma(x, x) = 0$ for all $x \in V(d,
q)$; such a space exists only if $d=2N$, where $N$ is called its
rank. A~subspace of $V(d, q)$ is called totally isotropic if
$\sigma$ vanishes identically on it. $\mathcal{W}(2N-1,q)$ can then be
regarded as the space of totally isotropic subspaces of $V(d, q)$.
The maximal totally isotropic subspaces of $V(d, q)$, also called} {\it generators} {of $\mathcal{W}(2N-1,q)$, have all the same dimension $N-1$. In what follows we shall only be concerned with $\mathcal{W}(2N-1,2)$; this space features} $|{\rm PG}(2N-1,2)|=2^{2N}-1=4^N-1$ points and the number of its generators amounts to $(2+1)(2^2+1)\cdots(2^N+1)$.

The generalized real $N$-qubit Pauli group, denoted by ${\cal P}_N$, is generated by $N$-fold tensor products of the matrices
\begin{gather*}
I = \left(
\begin{matrix}
1 & 0 \\
0 & 1
\end{matrix}
\right),\qquad
X = \left(
\begin{matrix}
0 & 1 \\
1 & 0
\end{matrix}
\right),~
Y = \left(
\begin{matrix}
0 & -1 \\
1 & 0
\end{matrix}
\right)
\qquad {\rm and}\qquad
Z = \left(
\begin{matrix}
1 & 0 \\
0 & -1
\end{matrix}
\right).
\end{gather*}
Explicitly,
\begin{gather*}
{\cal P}_N = \{\pm A_1 \otimes A_2 \otimes\cdots\otimes A_N: \, A_i \in \{I, X, Y, Z \},\; i = 1, 2,\dots,N \}.
\end{gather*}
The associated factor group $\overline{{\cal P}}_N \equiv {\cal P}_N/{\cal Z}({\cal P}_N)$, where the center ${\cal Z}({\cal P}_N)$ consists of $\pm I_{(1)} \otimes I_{(2)} \otimes \cdots \otimes I_{(N)}$, features $4^N$ elements.
The elements of $\overline{{\cal P}}_N \backslash \{I_{(1)} \otimes I_{(2)} \otimes \cdots \otimes I_{(N)}\}$  can be bijectively identif\/ied with the same number of points of
$\mathcal{W}(2N-1, 2)$ in such a way that two commuting elements of the group will lie on the same totally isotropic line of this polar space.
If one selects a basis of $\mathcal{W}(2N-1,2)$ in which the  symplectic form $\sigma(x,y)$ is given by
\begin{gather}\label{sympl}
\sigma(x,y) =
(x_1 y_{N+1} - x_{N+1} y_1) + (x_2 y_{N+2} - x_{N+2} y_2) + \dots + (x_N y_{2N} - x_{2N} y_N),
\end{gather}
then this bijection acquires the form:
\begin{gather}\label{corr1}
A_i \leftrightarrow (x_i, x_{i+N}), \qquad i \in \{1, 2, 3, 4\},
\end{gather}
with the understanding that
\begin{gather}\label{corr2}
I \leftrightarrow (0,0),\qquad X \leftrightarrow (0,1),\qquad Y \leftrightarrow (1,1),\qquad
Z \leftrightarrow (1,0);
\end{gather}
thus, for example, in $\mathcal{W}(7,2)$ the point having coordinates $(0,1,1,0,0,1,0,1)$ corresponds to the element $I \otimes Y \otimes Z \otimes X \equiv IYZX$.

The elements of the group $\overline{\mathcal{P}}_N$
whose square is $+I_{(1)}I_{(2)} \cdots I_{(N)}$ (i.e., symmetric elements)  lie on a certain  $\mathcal{Q}^{+}(2N - 1,2)$ of
the ambient space ${\rm PG}(2N-1, 2)$.
It follows from the def\/inition of the bijection that the equation of the $\mathcal{Q}^{+}(2N-1, 2)$ accommodating all symmetric elements must have the following standard form
\begin{gather}\label{hyperbolic}
\mathcal{Q}(x)=x_1x_{N+1} + x_2x_{N+2} + \dots + x_Nx_{2N} = 0.
\end{gather}
This can readily be inspected using the fact that the matrix $Y$ is the only skew-symmetric element in the set $\{I, X, Y, Z\}$  and, so, any symmetric element of the group must contain an even number of $Y$'s.

It should also be added that generators, of both $\mathcal{W}(2N - 1, 2)$ and $\mathcal{Q}^{+}(2N - 1, 2)$, correspond to {\it maximal} sets of mutually commuting elements of the group (see~\cite{hos} for a proof of this fact).

\section[Mapping ${\rm LGr}(N,2N)$ to ${\rm PG}(2^N-1,2)$]{Mapping $\boldsymbol{{\rm LGr}(N,2N)}$ to $\boldsymbol{{\rm PG}(2^N-1,2)}$}\label{bijection}

Recently, L\'evay, Planat and Saniga~\cite{PMM} found and analyzed in detail an explicit bijection between
the set of 135 maximum sets of mutually commuting elements of the three-qubit Pauli group
(that is, the set of generators of $\mathcal{W}(5,2)$) and the set of 135 symmetric operators of the four-qubit
Pauli group (that is, the set of points lying on a particular $\mathcal{Q}^{+}(7,2)$ of $\mathcal{W}(7,2)$).
Following the spirit of this work, we will generalize this physically important result and prove the existence of a similar bijection between {\it any} $N$-qubit and $2^{N-1}$-qubit Pauli groups. This will be done
by considering  f\/irst the Grassmaniann variety ${\rm Gr}(N,2N)$,
then its associated Lagrangian Grassmannian\footnote{Since (the def\/initions of) the two objects carry a lot of properties that are insensitive on the choice of the base f\/ield, our presentation will be following the classical case of the complex numbers~\cite{GKZ,L}.} ${\rm LGr}(N,2N)$ and, f\/inally, using a crucial fact that we work in characteristic $2$.

To this end in view, let us f\/irst recall the def\/inition of the variety of $N$-planes in $\KK^{2N}$, i.e.\ the Grassmannian variety ${\rm Gr}(N,2N)$.
An $N$-plane  (respectively an $(N-1)$-projective-plane) $P$, spanned by  $N$ non-zero vectors $u_1,u_2,\dots,u_N$ of $\KK^{2N}$
(respectively by  $N$ points $[u_1], [u_2],\dots,[u_N]$ of ${\rm PG}(2N-1,2)$) is a point of the Grassmannian variety ${\rm Gr}(N,2N)\subset  \PP(\wedge ^N \KK^{2N})={\rm PG}(\binom{2N}{N}-1,2)$.
The embedding of the Grassmannian variety is given by the so-called Pl\"ucker map:
\[
P=\text{span}\langle u_1,u_2,\dots,u_N \rangle \mapsto [u_1\wedge u_2\wedge \cdots\wedge u_N] \in {\rm Gr}(N,2N)\subset \PP\big(\wedge ^N \KK^{2N}\big).
\]
In other words, the Grassmanian variety is the set of all skew symmetric tensors that can be factorized (i.e., are separable).
The algebraic equations def\/ining ${\rm Gr}(N,2N)$ are known as the Pl\"ucker equations. Let $(e_i)_{1\leq i\leq 2N}$ be a basis of the
vector space $\KK^{2N}$ and let $P\in \PP(\wedge ^N \KK^{2N})$,~i.e.
\[
P=\sum_{1\leq i_1<i_2<\cdots<i_N\leq 2N} p_{i_1,i_2,\dots, i_N} e_{i_1}\wedge e_{i_2}\wedge \cdots \wedge e_{i_N}.
\]
If $P$ belongs to ${\rm Gr}(N,2N)$, then for any two sequences $1\leq i_1<\dots<i_{N-1}\leq 2N$ and $1\leq j_1<\cdots <j_{N+1}\leq 2N$, the coordinates of~$P$ satisfy the following
relations (see \cite[p.~94]{GKZ})
\begin{gather}\label{pluckerequation}
 \sum_{a=1}^{N+1} (-1)^a p_{i_1,i_2,\dots,i_{N-1},j_a}p_{j_1,j_2,\dots,\hat{j_a},\dots,j_{N+1}}=0,
\end{gather}
where the symbol $\hat{j_a}$ means that the corresponding index is omitted.
Equivalently, the coordinates $[p_{1,2,\dots,N},\dots,p_{N,N+1\dots,2N}]$ of $P\in {\rm Gr}(N,2N)$ can be expressed as follow. Let $M$ be an $N\times 2N$
matrix whose rows are coordinates of $N$ vectors that are spanning the $N$-plane $P$. Then, we have
\begin{gather*}
p_{i_1,\dots,i_N}=\Delta_{\{i_1,\dots,i_N\},\{1,\dots,N\}}(M).
\end{gather*}
We are only interested  in those $(N-1)$-planes of ${\rm PG}(2N-1,2)$ which are totally isotropic with respect to our symplectic form $\sigma$ (i.e., in generators of $\mathcal{W}(2N-1,2)$).
The extension of~$\sigma$ to $\PP(\wedge^N \KK^{2N})$ def\/ines (see~\cite{CZ}) linear conditions  on the coordinates
$[p_{1,\dots,N},p_{1,\dots,N-1,N+1},\dots$, $p_{N,\dots,2N}]$ of~$P$
to insure that $P$ is totally isotropic. These linear conditions  def\/ine a projective space~$\PP(L)$ whose intersection with ${\rm Gr}(N,2N)$
is a sub-variety of ${\rm Gr}(N,2N)$ called the Lagrangian variety,
\[
{\rm LGr}(N,2N)={\rm Gr}(N,2N)\cap \PP(L).
\]
The Lagrangian variety is thus the variety of all the generators ${\rm PG}(N-1,2)$
of $\mathcal{W}(2N-1,2)$.
We will now show that over $\KK$ the variety ${\rm LGr}(N,2N)$ can further be projected bijectively to a subset of points of ${\rm PG}(2^N-1,2)$,
where ${\rm PG}(2^N-1,2)$ is the projective space obtained by eliminating the variables involved in the linear conditions which
def\/ine $\PP(L)$ (i.e., the linear conditions
given by the extension of $\sigma$ to $\PP(\wedge ^N \KK^{2N})$).

Let
\begin{gather*}
P=\bigg(\!e_1+\sum_j a_{1,j}e_{N+j}\!\bigg)\!\wedge \!\bigg(\!e_2+\sum_j a_{2,j}e_{N+j}\!\bigg)\!\wedge \cdots \wedge\! \bigg(\!e_N+\sum_j a_{N,j}e_{N+j}\!\bigg) \! \in {\rm Gr}(N,2N).
\end{gather*}
Expanding this expression, we obtain the local parametrization of ${\rm Gr}(N,2N)$:
\begin{gather*}
P=e_1\wedge\dots\wedge e_N+\sum_{i=1}^N\sum_{j=1}^N a_{ij} e_1\wedge\dots\wedge e_{i-1}\wedge e_{N+j}\wedge e_{i+1}\wedge e_{N}\\
\hphantom{P=}{} +
\sum_{i,j}\sum_{s,t} (a_{is}a_{jt}-a_{it}a_{js})e_1\wedge\dots \wedge e_{i-1}\wedge e_{N+s}\wedge e_{i+1}\wedge \cdots \\
\hphantom{P=}{} \cdots
\wedge e_{j-1}\wedge e_{N+t}\wedge e_{j+1}\wedge \dots\wedge e_N+\cdots.
\end{gather*}
This shows that locally the coordinates of $P$ can be written as
\begin{gather}\label{symmetric}
[1,a_{11},\dots,a_{NN},a_{11}a_{22}-a_{21}a_{12},\dots]=[1,\Delta_{1}(A),\dots,\Delta_{I,J}(A),\dots],
 \end{gather}
where $A=(a_{i,j})$.
Requiring $P$ to be totally isotropic means that the vectors spanning $P$
 must be totally isotropic. Denoting $u_i=e_i+\sum_j a_{i,j}e_{N+j}$,
we get $\sigma(u_s,u_t)=a_{st}-a_{ts}$, which is zero if and only if $A=(a_{ij})$ is a symmetric matrix. Thus $P$ will be totally isotropic if its
coordinates locally  correspond to minors of a symmetric matrix $A$ over $\KK$.

The linear conditions def\/ining $\PP(L)$ correspond locally to the fact that
the minors $\Delta_{I,J}(A)$ and $\Delta_{J,I}(A)$ are equal for $I\neq J$. Moreover, these  conditions
do not involve the coordinates corresponding to principal minors. Thus,  we obtain a splitting $\wedge^N \KK^{2N}=V+W$ such that
the coordinates def\/ining~$V$ are locally given by minors of type $\Delta_{I,J}$, whereas the coordinates def\/ining~$W$ are principal minors~$\Delta_I (A)$.
But for symmetric matrices over $\KK={\rm GF}(2)$ all of\/f-diagonal entries are completely determined by the principal minors $\Delta_{\{i\}}(A)$ and $\Delta_{\{i,j\}}(A)$. This fact readily stems from the following equation
$a_{ii}a_{jj}-a_{ij}^2=\Delta_{\{i,j\}}(A)$, i.e., $a_{i,j}^2=\Delta_i(A)\Delta_j(A)-\Delta_{i,j}(A)$. Thus, all minors $\Delta_{I,J}(A)$, with $I\neq J$, are (over $\KK$)
{\it uniquely} determined
by the principal minors $\Delta_{K}(A)$ of $A$. In other words, once the coordinates in $W$ of a~point of ${\rm LGr}(N,2N)$ are chosen, the coordinates
in~$V$ are automatically f\/ixed. If we consider the cone $\widehat{{\rm LGr}}(N,2N)\subset \wedge^N \KK^{2N}=V+W$, this can be regarded as a~graph over mere $W$ and mapped
bijectively to a~subset of~$W$. The dimension of~$W$ is given by the number of principal minors: $\sum\limits_{i=0} ^N \binom{N}{i}=2^N$. Since all principal minors cannot vanish simultaneously, we obtain a~well-def\/ined projective map $\underline{\pi}:{\rm LGr}(N,2N)\to \PP(W)={\rm PG}(2^N-1,2)$.
The map~$\underline{\pi}$ sends $P$ to $p\in \PP(W)$, where $p$ is def\/ined by the coordinates of~$P$ not occurring in the equations def\/ining~$V$.
All in all, we obtain a~bijective mapping by projecting~${\rm LGr}(N,2N)$ to~${\rm PG}(2^N-1,2)$ after eliminating all the variables involved/occurring
in the linear conditions.

\section{An explicit construction of the bijection: a few examples}\label{example}

The above-given proof of the existence of the mapping
\[
\underline{\pi}: \ {\rm LGr}(N,2N)\to {\rm PG}\big(2^{N}-1,2\big)
\]
provides us with a recipe of how to obtain the equations
of the image \[
\underline{\pi}({\rm LGr}(N,2N)\subset {\rm PG}\big(2^N-1,2\big).
\] Indeed, following our reasoning one f\/irst needs to f\/ind the ideal $I({\rm Gr}(N,2N))$
(i.e., a set of equations) def\/ining
${\rm Gr}(N,2N)$, as well as the linear conditions $J=(l_1,\dots,l_m)$ induced by the associated symplectic form. These two sets of equations will then def\/ine the ideal of ${\rm LGr}(N,2N)$,
i.e., \[
I({\rm LGr}(N,2N))=I({\rm Gr}(N,2N))\cup J.
\]
Then we calculate the ideal of the projection $\underline{\pi}({\rm LGr}(N,2N))$ by eliminating in $I({\rm LGr}(N,2N))$ all the variables appearing in $J$.
The last step can be done by hand when cases are rather simple, or be handled with the formalism of the elimination theory \cite{CLO} when calculations become more tedious.
This formalisms provides algorithms to compute the ideal of the projection (more precisely, the ideal $I$ such that the variety $V(I)$ contains the projection). In practice, however, with increasing $N$ we quickly face insurmountable computational dif\/f\/iculties, as explicitly pointed out at the end of this section.
We shall now illustrate this approach on the f\/irst three cases in the sequence.

\subsection[The smallest non-trivial $({\rm Gr}(2,4) \mapsto {\rm LGr}(2,4)\simeq {\rm PG}(3,2))$ case]{The smallest non-trivial $\boldsymbol{({\rm Gr}(2,4) \mapsto {\rm LGr}(2,4)\simeq {\rm PG}(3,2))}$ case}

The set of lines in ${\rm PG}(3,2)$ (or $2$-planes in $\KK^4$) is the f\/irst non-trivial Grassmannian variety.
From equations~(\ref{pluckerequation}) it follows that ${\rm Gr}(2,4)$ is def\/ined by a single equation,
\begin{gather*}
 p_{12}p_{34}-p_{13}p_{24}+p_{14}p_{23}=0,
\end{gather*}
representing a quadric surface in ${\rm PG}(\binom{4}{2}-1,2)={\rm PG}(5,2)$.
Hence, ${\rm Gr}(2,4)$ is a variety of dimension $4$.
The canonical symplectic form (see equation~(\ref{sympl})) $\sigma(x,y)=x_1y_3-x_3y_1+x_2y_4-x_4y_2$ imposes that
$P$ (of projective coordinates $[p_{12}:p_{13}:p_{14}:p_{34}:p_{24}:p_{23}]$) is isotropic if and only if $p_{13}=p_{24}$. The linear
conditions stemming from $\sigma$ tell
us that
\[
{\rm LGr}(2,4)={\rm Gr}(2,4)\cap {\rm PG}(4,2),
\] where ${\rm PG}(4,2)=\{[x_1:x_2:x_3:x_4:x_5:x_6]\in {\rm PG}(5,2), x_2=x_5\}$. The variety ${\rm LGr}(2,4)$ is of dimension $3$, being mapped down to ${\rm PG}(3,2)$
when we take into account the projection $\underline{\pi}:{\rm LGr}(2,4)\to {\rm PG}(3,2)$ given by $\underline{\pi}([x_1:x_2:x_3:x_4:x_5:x_6])=[x_1:x_3:x_4:x_6]$.
It is a~one-to-one mapping (because both $p_{13}$ and $p_{24}$ are f\/ixed by the other minors), so we have
\[
\underline{\pi}({\rm LGr}(2,4))={\rm PG}(3,2).
\]

\subsection[The `L\'evay--Planat--Saniga' $({\rm Gr}(3,6) \mapsto {\rm LGr}(3,6) \mapsto \mathcal{Q}^{+}(7, 2))$ case revisited]{The `L\'evay--Planat--Saniga' $\boldsymbol{({\rm Gr}(3,6) \mapsto {\rm LGr}(3,6) \mapsto \mathcal{Q}^{+}(7, 2))}$ case\\ revisited}\label{examples3}

For $N=3$,  equations~(\ref{pluckerequation}) reduce into the following form
\begin{gather*}
\sum_{a=1}^{4} (-1)^a p_{i_1 i_2 j_a} p_{j_1 \dots \widehat{j}_a \dots j_{4}} = 0.
\end{gather*}
For each choice of the pair of indices of $\{i_1, i_2\}$ we f\/ind eight equations with three terms each and one equation featuring all four terms. There will be, of course, an overlap and what we
get are only 30 independent three-term equations and f\/ive four-term ones. We start with   $\{i_1, i_2\} = \{1,2\}$ and for each
subsequent choice of these two indices we list only those equations that have not appeared in the preceding steps. Each equation of the former set is, apart from the pair~$i_1$,~$i_2$, characterized by the string $\{j_1, j_2, j_3, j_4\}$ shown after the equation. The 30 equations read:
\begin{gather*}
\{i_1, i_2\} = \{1,2\}: \\
p_{123} p_{145} + p_{124} p_{135} + p_{125} p_{134} = 0,\qquad \{1,3,4,5\},\\
p_{123} p_{146} + p_{124} p_{136} + p_{126} p_{134} = 0,\qquad \{1,3,4,6\},\\
p_{123} p_{156} + p_{125} p_{136} + p_{126} p_{135} = 0,\qquad \{1,3,5,6\},\\
p_{124} p_{156} + p_{125} p_{146} + p_{126} p_{145} = 0,\qquad \{1,4,5,6\},\\
p_{123} p_{245} + p_{124} p_{235} + p_{125} p_{234} = 0,\qquad \{2,3,4,5\},\\
p_{123} p_{246} + p_{124} p_{236} + p_{126} p_{234} = 0,\qquad \{2,3,4,6\},\\
p_{123} p_{256} + p_{125} p_{236} + p_{126} p_{235} = 0,\qquad \{2,3,5,6\},\\
p_{124} p_{256} + p_{125} p_{246} + p_{126} p_{245} = 0,\qquad \{2,4,5,6\}, \\
\{i_1, i_2\} = \{1,3\}:\\
p_{134} p_{156} + p_{135} p_{146} + p_{136} p_{145} = 0,\qquad \{1,4,5,6\},\\
 p_{123} p_{345} + p_{134} p_{235} + p_{135} p_{234} = 0,\qquad \{2,3,4,5\},\\
 p_{123} p_{346} + p_{134} p_{236} + p_{136} p_{234} = 0,\qquad \{2,3,4,6\},\\
 p_{123} p_{356} + p_{135} p_{236} + p_{136} p_{235} = 0,\qquad \{2,3,5,6\},\\
p_{134} p_{356} + p_{135} p_{346} + p_{136} p_{345} = 0,\qquad \{3,4,5,6\},
\\
\{i_1, i_2\} = \{1,4\}:  \\
p_{124} p_{345} + p_{134} p_{245} + p_{145} p_{234} = 0,\qquad \{2,3,4,5\},\\
p_{124} p_{346} + p_{134} p_{246} + p_{146} p_{234} = 0,\qquad \{2,3,4,6\},\\
p_{124} p_{456} + p_{145} p_{246} + p_{146} p_{245} = 0,\qquad \{2,4,5,6\},\\
p_{134} p_{456} + p_{145} p_{346} + p_{146} p_{345} = 0,\qquad \{3,4,5,6\},\\
\{i_1, i_2\} = \{1,5\}:\\
p_{125} p_{345} + p_{135} p_{245} + p_{145} p_{235} = 0,\qquad \{2,3,4,5\},\\
p_{125} p_{356} + p_{135} p_{256} + p_{156} p_{235} = 0,\qquad \{2,3,5,6\},\\
p_{125} p_{456} + p_{145} p_{256} + p_{156} p_{245} = 0, \qquad \{2,4,5,6\},\\
p_{135} p_{456} + p_{145} p_{356} + p_{156} p_{345} = 0, \qquad \{3,4,5,6\},\\
\{i_1, i_2\} = \{1,6\}: \\
p_{126} p_{346} + p_{136} p_{246} + p_{146} p_{236} = 0,\qquad \{2,3,4,6\},  \\
p_{126} p_{356} + p_{136} p_{256} + p_{156} p_{236} = 0,\qquad \{2,3,5,6\},  \\
p_{126} p_{456} + p_{146} p_{256} + p_{156} p_{246} = 0,\qquad \{2,4,5,6\},   \\
p_{136} p_{456} + p_{146} p_{356} + p_{156} p_{346} = 0,\qquad \{3,4,5,6\},  \\
\{i_1, i_2\} = \{2,3\}:\\
p_{234} p_{256} + p_{235} p_{246} + p_{236} p_{245} = 0,\qquad \{2,4,5,6\},\\
p_{234} p_{356} + p_{235} p_{346} + p_{236} p_{345} = 0,\qquad \{2,4,5,6\},\\
\{i_1, i_2\} = \{2,4\}:\\
p_{234} p_{456} + p_{245} p_{346} + p_{246} p_{345} = 0,\qquad \{3,4,5,6\},\\
\{i_1, i_2\} = \{2,5\}:  \\
p_{235} p_{456} + p_{245} p_{356} + p_{256} p_{345} = 0,\qquad  \{3,4,5,6\}, \\
\{i_1, i_2\} = \{2,6\}:\\
  p_{236} p_{456} + p_{246} p_{356} + p_{256} p_{346} = 0,\qquad \{3,4,5,6\}.
\end{gather*}
The f\/ive independent four-term equations  (followed by the corresponding pair $\{i_1, i_2\}$) are
 \begin{gather*}
p_{123} p_{456} + p_{124} p_{356} + p_{125} p_{346} + p_{126} p_{345} = 0,\qquad \{1,2\},\\
p_{123} p_{456} + p_{134} p_{256} + p_{135} p_{246} + p_{136} p_{245} = 0,\qquad \{1,3\},\\
p_{124} p_{356} + p_{134} p_{256} + p_{145} p_{236} + p_{146} p_{235} = 0,\qquad \{1,4\},\\
p_{125} p_{346} + p_{135} p_{246} + p_{145} p_{236} + p_{156} p_{234} = 0,\qquad \{1,5\},\\
p_{126} p_{345} + p_{136} p_{245} + p_{146} p_{235} + p_{156} p_{234} = 0,\qquad \{1,6\}.
\end{gather*}
We are only interested in the Lagrangian grassmannian ${\rm LGr}(3,6)$, that is in those planes of ${\rm PG}(5,2)$ that are totally isotropic with respect to a given symplectic polarity.
Choosing the latter to have again the `canonical' form (equation~(\ref{sympl})),
\begin{gather*}
(x_1 y_4 - x_4 y_1) + (x_2 y_5 - x_5 y_2) + (x_3 y_6 - x_6 y_3) = 0,
\end{gather*}
the coordinates of such planes have to meet the following constraints
\begin{gather}
p_{125} = p_{136}, \qquad p_{235} = p_{134},\qquad  p_{124} = p_{236}, \nonumber\\
 p_{245} = p_{346}, \qquad p_{256} = p_{146}, \qquad p_{145} = p_{356},\label{constr}
\end{gather}
which reduce the set of 30 three-term equations into
\begin{alignat*}{3}
& p_{123} p_{356} + p_{236} p_{135} + p_{136} p_{235} = 0, \qquad && p_{236} p_{456} + p_{356} p_{246} + p_{256} p_{346} = 0, &\\
& p_{123} p_{256} + p_{236} p_{136} + p_{126} p_{235} = 0, \qquad && p_{235} p_{456} + p_{356} p_{346} + p_{256} p_{345} = 0, &\\
& p_{236} p_{156} + p_{136} p_{256} + p_{126} p_{356} = 0, \qquad && p_{136} p_{456} + p_{356} p_{256} + p_{156} p_{346} = 0, &\\
& p_{123} p_{346} + p_{236} p_{235} + p_{136} p_{234} = 0, \qquad && p_{123} p_{156} + p_{136}^{2} + p_{126} p_{135} = 0, & \\
& p_{236} p_{256} + p_{136} p_{246} + p_{126} p_{346} = 0, \qquad && p_{123} p_{246} + p_{236}^{2} + p_{126} p_{234} = 0, &\\
& p_{235} p_{156} + p_{135} p_{256} + p_{136} p_{356} = 0, \qquad && p_{123} p_{345} + p_{235}^{2} + p_{135} p_{234} = 0, &\\
& p_{235} p_{356} + p_{135} p_{346} + p_{136} p_{345} = 0, \qquad && p_{135} p_{456} + p_{356}^{2} + p_{156} p_{345} = 0, &\\
& p_{236} p_{345} + p_{235} p_{346} + p_{356} p_{234} = 0, \qquad && p_{126} p_{456} + p_{256}^{2} + p_{156} p_{246} = 0, &\\
& p_{236} p_{346} + p_{235} p_{246} + p_{256} p_{234} = 0, \qquad && p_{234} p_{456} + p_{346}^{2} + p_{246} p_{345} = 0,&
\end{alignat*}
and the set of f\/ive four-term ones into
\begin{gather*}
p_{123} p_{456} + p_{236} p_{356} + p_{136} p_{346} + p_{126} p_{345} = 0, \\
p_{123} p_{456} + p_{235} p_{256} + p_{135} p_{246} + p_{136} p_{346} = 0, \\
p_{136} p_{346} + p_{135} p_{246} + p_{356} p_{236} + p_{156} p_{234} = 0, \\
p_{126} p_{345} + p_{136} p_{346} + p_{256} p_{235} + p_{156} p_{234} = 0.
\end{gather*}
These last four equations are, however, not independent, as each of them is equal to the sum of the remaining three. Moreover, summing the f\/irst of them with the third one, or the second with the fourth, yields
\begin{gather}\label{quadric36}
p_{123} p_{456} + p_{126} p_{345} + p_{135} p_{246} + p_{156} p_{234} = 0,
\end{gather}
which after relabeling the variables as $x_1=p_{123}$, $x_2=p_{126}$, $x_3=p_{135}$, $x_4=p_{156}$, $x_5=p_{456}$, $x_6=p_{345}$, $x_7=p_{246}$ and $x_8=p_{234}$ reads
\begin{gather*}
 x_1x_5+x_2x_6+x_3x_7+x_4x_8=0
\end{gather*}
and is readily recognized to represent a hyperbolic quadric $\mathcal{Q}^{+}(7, 2)$ in a particular subspace ${\rm PG}(7,2)$ of the ambient projective space ${\rm PG}(19,2)$ of ${\rm Gr}(3,6)$.
More precisely, the quadric def\/ined by equation~(\ref{quadric36}) lives in the ideal $I_2({\rm LGr}(3,6))$ and it is the {\it only} quadric that does not depend on the coordinates
$p_{136}$, $p_{236}$, $p_{235}$, $p_{356}$, $p_{256}$, and $p_{346}$.
Thus, if we consider the splitting of the linear space $\KK^{14}=\KK^8 \oplus \KK^6$, where $\KK^8$ represents the vector space def\/ined by
the set of coordinates $\{p_{123}$, $p_{126}$, $p_{135}$, $p_{156}$, $p_{456}$, $p_{345}$, $p_{246}$, $p_{234}\}$ and $\KK^6$ that def\/ined by
 $\{p_{136}$, $p_{236}$, $p_{235}$, $p_{356}$, $p_{256}$, $p_{346}\}$, and employ the fact that each coordinate
from the latter set can be expressed as a linear combination of the coordinates from the former set, then we can
represent $\widehat{{\rm LGr}}(3,6)\subset \KK^{14}$
as a graph over the quadric $\mathcal{Q}^{+}(7, 2)$ def\/ined by equation~(\ref{quadric36}) in $\KK^8$, i.e.,
\[
\widehat{{\rm LGr}}(3,6)=\big\{({\bf x},g({\bf x}))\in \KK^{14}, {\bf x}\in \widehat{\mathcal{Q}}^{+} (7,2)\subset \KK^8\big\}.
\]
One thus automatically gets a bijection between ${\rm LGr}(3,6)$ and the $\mathcal{Q}^+(7,2)$ by
taking the projection to the base of the graph $\KK^8\oplus \KK^6\to \KK^8$.

This procedure  can be rephrased in algebraic terms in the framework of elimination theory~\cite{CLO}.
Given an ideal $I\subset \KK[x_1,\dots,x_n]$, the $l$-th elimination ideal, with $1\leq l \leq n$, is the
ideal of $\KK[x_{l+1},\dots,x_n]$ def\/ined by
\[
I_l=I\cap \KK[x_{l+1},\dots,x_n].
\]
Let $\pi_l$ be the projection $\KK^n\to \KK^{n-l}$ def\/ined by $\pi_l(a_1,\dots,a_n)=(a_{l+1},\dots,a_n)$.
If $V(I)=\{(a_1,\dots,a_n)\in \KK^n, f(a_1,\dots,a_n)=0, \forall\, f\in I\}$ is an af\/f\/ine variety
corresponding to the ideal~$I$, then $\pi(V(I))\subset V(I_l)$, i.e., the projection of~$V(I)$ is
contained in the algebraic variety def\/ined by the elimination ideal (which is, in fact, the smallest af\/f\/ine variety containing $\pi(V)$).
Using the notion of Groebner basis, one can compute the elimination ideal $I$ from the fact that
\[
G_l=G\cap \KK[x_{l+1},\dots,x_n],
\]
where~$G$ is the Groebner basis of~$I$ and~$G_l$ that of~$I_l$. To perform this calculation, it
suf\/f\/ices to choose a monomial order adapted to eliminate the f\/irst~$l$ variables.

Going back to the (cone over the) Lagrangian Grassmanian, $\widehat{{\rm LGr}}(3,6)\subset \KK^{14}$, we know that the ideal of this variety
is def\/ined by the above-given $21$ equations
of degree $2$. This ideal is obtained from the Pl\"ucker relations def\/ining ${\rm Gr}(3,6)$ by imposing the six constraints (\ref{constr}), i.e.
\[
I({\rm LGr}(3,6))=I({\rm Gr}(3,6)\cup J,
\]
where $J=(p_{125}+p_{136},p_{235}+p_{134},p_{124}+p_{236},p_{245}+p_{346},p_{256}+p_{146},p_{145}+p_{356})$.
The Pl\"ucker coordinates appearing in $J$ are exactly those we want to eliminate to project down to $\KK^8$, because they depend linearly on the remaining ones.
Starting form the ideal $I(\widehat{{\rm LGr}}(3,6))$, we compute the desired elimination ideal using Macaulay2:
\begin{gather*}
\begin{split}
&I(\widehat{{\rm LGr}}(3,6))\cap \KK[p_{123},p_{126},p_{135},p_{156},p_{456},p_{345},p_{246},p_{234}]\\
& \qquad {} =(p_{123}p_{456}+p_{126}p_{345}+p_{135}p_{246}+p_{156}p_{234}).
\end{split}
\end{gather*}
Thus, $\pi(\widehat{{\rm LGr}}(3,6))\subset \widehat{\mathcal{Q}}^+(7,2)$. Everything holds projectively, as we worked only
with homogeneous polynomials; then $\underline{\pi}({\rm LGr}(3,6))\subset \mathcal{Q}^+(7,2)$, where $\underline{\pi}: {\rm PG}(13,2)\backslash {\rm PG}(5,2) \to {\rm PG}(7,2)$.\footnote{It is worth mentioning here that the  projection map is  well def\/ined only outside ${\rm PG}(5,2)=\{p\in {\rm PG}(13,2), p=[a_1,\dots,a_6,0,\dots,0]\}$; however, $\underline{\pi}|_{{\rm LGr}(3,6)}$ is well def\/ined because the variables $p_{123},\dots,p_{234}$  do not vanish simultaneously on ${\rm LGr}(3,6)$.}
Moreover, since both $\sharp \underline{\pi}({\rm LGr}(3,6))=135$ and $\sharp \mathcal{Q}^+(7,2)=135$ and because
$\underline{\pi}$ is a bijection (the values of $p_{125},\dots,p_{356}$ are completely determined by those of $p_{123},\dots,p_{234}$),
we f\/inally get
\[
\underline{\pi}({\rm LGr}(3,6))=\mathcal{Q}^+(7,2).
\]

\subsection[A more intricate $({\rm Gr}(4,8) \mapsto {\rm LGr}(4,8) \mapsto \mathcal{Q}^{+}(15, 2))$ case]{A more intricate $\boldsymbol{({\rm Gr}(4,8) \mapsto {\rm LGr}(4,8) \mapsto \mathcal{Q}^{+}(15, 2))}$ case}\label{examples4}

We shall follow the same strategy to show that the 2295 maximal subspaces of $\mathcal{W}(7,2)$ are mapped to a subset of points of a hyperbolic quadric
$\mathcal{Q}^{+}(15, 2)$.
${\rm Gr}(4,8)$ is a variety of the ${8 \choose 4} - 1 = 69$-dimensional projective space def\/ined by
\begin{gather*}
\sum_{a=1}^{5} (-1)^a p_{i_1 i_2 i_3 j_a} p_{j_1 \dots \widehat{j}_a \dots j_{5}} = 0.
\end{gather*}
The Lagrangian Grassmannian associated with the following symplectic polarity
\[(x_1 y_5 - x_5 y_1) + (x_2 y_6 - x_6 y_2) + (x_3 y_7 - x_7 y_3) + (x_4 y_8 - x_8 y_4) = 0,\]
has to meet 24 constraints of the type
 \begin{gather*}
p_{1345} = p_{2346},\qquad p_{1245} = p_{2347},\qquad p_{1235}=p_{2348},\qquad
p_{1246}=p_{1347},\qquad p_{1236}=p_{1348},\\
 p_{1237}=p_{1248}, \qquad
p_{1358} = p_{2368},\qquad p_{1258} = p_{2378},\qquad p_{1257}=p_{2478},\qquad
p_{1268}=p_{1378},\\
 p_{1267}=p_{1478},\qquad p_{1367}=p_{1468}, \qquad
p_{1457} = p_{2467},\qquad p_{1456} = p_{3467},\qquad p_{1356}=p_{3468},\\
p_{2456}=p_{3457},\qquad p_{2356}=p_{3458},\qquad p_{2357}=p_{2458}, \qquad
p_{1578} = p_{2678},\qquad p_{1568} = p_{3678},\\ p_{1567}=p_{4678},\qquad
p_{2568}=p_{3578},\qquad p_{2567}=p_{4578},\qquad p_{3567}=p_{4568},
\end{gather*}
and three conditions of the type
 \begin{gather*}
p_{1256} + p_{1357} + p_{1458} = 0, \qquad
p_{1256} + p_{2367} + p_{2468} = 0, \qquad
p_{1357} + p_{2367} + p_{3478} = 0.
\end{gather*}
We thus f\/ind 27 independent linear relations; hence, our ${\rm LGr}(4,8)$ lives in a subspace of the ${\rm PG}(69,2)$ that is isomorphic to ${\rm PG}(42,2)$.

To f\/ind the projection $\pi:\KK^{70} \to \KK^{16}$ we compute the elimination ideal
\[
I=(I({\rm Gr}(4,8))\cup J)\cap\KK[x_1,\dots,x_{16}],
\]
where $J$ is a homogeneous ideal of degree one generated by the above-given 27 equations
and $x_i$'s, $i \in \{1,2,\dots,16\}$, stand for the following $16$ Pl\"ucker coordinates that do not appear in the expression def\/ining $J$:
 \begin{alignat*}{5}
& x_1=p_{1234},\qquad && x_2=p_{1238},\qquad && x_3=p_{1247},\qquad && x_4=p_{1278},& \\
& x_5=p_{1346},\qquad && x_6=p_{1368},\qquad && x_7=p_{1467},\qquad && x_8=p_{1678}, & \\
& x_9=p_{5678},\qquad && x_{10}=p_{4567},\qquad && x_{11}=p_{3568},\qquad && x_{12}=p_{3456}, & \\
& x_{13}=p_{2578},\qquad && x_{14}=p_{2457},\qquad && x_{15}=p_{2358},\qquad && x_{16}=p_{2345}. &
\end{alignat*}
Using Macaulay2, we get the ideal
\begin{gather}\label{ideal}
I=(\mathcal{Q}_1,\mathcal{Q}_2,\mathcal{Q}_3,\mathcal{Q}_4,\mathcal{Q}_5,\mathcal{Q}_6,\mathcal{Q}_7,\mathcal{Q}_8,\mathcal{Q}_9,\mathcal{Q}_{10}),
\end{gather}
where $\mathcal{Q}_i$ are the following quadratic forms
 \begin{alignat*}{3}
& \mathcal{Q}_1=x_{12}x_{13} + x_{11}x_{14} + x_{10}x_{15} + x_{9}x_{16}, \qquad &&
\mathcal{Q}_2=x_{1}x_{13} + x_{2}x_{14} + x_{3}x_{15}+ x_{4}x_{16},& \\
& \mathcal{Q}_3=x_{1}x_{11} + x_{2}x_{12} + x_{5}x_{15} + x_{6}x_{16},\qquad &&
\mathcal{Q}_4=x_{4}x_{5} + x_{3}x_{6} + x_{2}x_{7} + x_{1}x_{8},&\\
& \mathcal{Q}_5=x_{1}x_{10} + x_{3}x_{12} + x_{5}x_{14} + x_{7}x_{16},\qquad &&
\mathcal{Q}_6=x_{5}x_{9} +x_{6}x_{10} + x_{7}x_{11} + x_{8}x_{12},& \\
& \mathcal{Q}_7=x_{3}x_{9} + x_{4}x_{10} + x_{7}x_{13} + x_{8}x_{14},\qquad &&
  \mathcal{Q}_8=x_{2}x_{9} + x_{4}x_{11} + x_{6}x_{13} + x_{8}x_{15},& \\
& \mathcal{Q}_9=x_{1}x_{9} + x_{4}x_{12} + x_{6}x_{14} +x_{7}x_{15},\qquad &&
\mathcal{Q}_{10}=x_{2}x_{10} + x_{3}x_{11} +x_{5}x_{13} + x_{8}x_{16}.&
\end{alignat*}
Moreover, the quadric
\[
\mathcal{Q}_0 \equiv \mathcal{Q}_9+\mathcal{Q}_{10} = 0
\]
is a hyperbolic quadric in ${\rm PG}(15,2)$ def\/ined by equation~(\ref{hyperbolic}).
The ideal $I$ def\/ines a subvariety~$V(I)$ of ${\rm PG}(15,2)$ that ${\rm LGr}(4,8)$ is mapped to. Moreover, $V(I)\subset \mathcal{Q}^+(15,2)$ because $\mathcal{Q}_0 \in I$. Again, by comparing
the number of points of $V(I)$ and ${\rm LGr}(4,8)$, we conclude that
\[
\underline{\pi}({\rm LGr}(4,8))=V(I) \subset \mathcal{Q}^+(15,2).
\]
{In both examples ($N=3$ and $N=4$) the set of equations obtained provides not only a set of equations which cuts out the variety}
$\underline{\pi}({\rm LGr}(N,2N))$,
 but also the ideal $I$ of the variety. This is obvious for $N=3$, because the ideal is principal and generated by an irreducible polynomial.
For $N=4$ we can directly check with Macaulay2 that the ideal $I$ of equation~(\ref{ideal}) {is prime, being thus also the ideal of }$\underline{\pi}({\rm LGr}(4,8))$. We shall return to this point in Section~\ref{variety_minor}.
As~$N$ increases, the calculations are more and more time-consuming and put also big demand on memory resources. The case $N=5$ was already out of reach for our computers.

\section[Stratification of ${\rm PG}(3,2)$, ${\rm PG}(7,2)$ and ${\rm PG}(15,2)$]{Stratif\/ication of $\boldsymbol{{\rm PG}(3,2)}$, $\boldsymbol{{\rm PG}(7,2)}$ and $\boldsymbol{{\rm PG}(15,2)}$}\label{partition}

Let us consider the natural action of the group $G={\rm SL}(2,2)\times {\rm SL}(2,2)\times\cdots \times {\rm SL}(2,2)\rtimes S_N$ on ${\rm PG}(2^N-1,2)$.
This action partitions the set of points of the projective space in terms of $G$-orbits and allows us, thanks to the bijection described in Section~\ref{bijection}, to partition
${\rm LGr}(N,2N)$ in terms of $G$-equivalent classes. Knowing a representative of each orbit, we can use the equations obtained in the previous section
to check whether a particular orbit does or does not belong to $\underline{\pi}({\rm LGr}(N,2N))$.
The $G$-action preserves also dif\/ferent notions of rank. First, it preserves the tensor rank ($T$-rank), which can be def\/ined as follows:
if $p\in {\rm PG}(2^N-1,2)$ we will say
that~$p$ has $T$-rank $k$ if $p=\Big[\sum\limits_{i=1} ^k v_{1}^i\otimes v_2^i\otimes\cdots\otimes v_n^i\Big]$, where $v_j^i \in (\KK^2)_i$
and $k$ is the minimal integer such that this property holds  (see~\cite{aft, bs}  for recent work on tensor rank over ${\rm GF}(2)$).
Another notion of rank, more interesting in our situation, is that of exclusive rank, $E$-rank, as proposed in~\cite{O3}.
{The $E$-rank is only def\/ined for points $p$ of} $\underline{\pi}({\rm LGr}(N,2N))$.  Given  $p\in \underline{\pi}({\rm LGr}(N,2N))$
there exists a unique $P\in {\rm LGr}(N,2N)$ which is in local coordinates def\/ined by a symmetric matrix $A$ (Section~\ref{bijection} and equation~(\ref{symmetric})). Then we will say that
$p$ is of $E$-rank $k$ if, and only if, all $(k+1)\times (k+1)$ exclusive minors of $A$, i.e.\ minors $\Delta_{I,J}(A)$ with $I\cap J=\varnothing$, are zero.

In the following examples, we use the classif\/ication of $G$-orbits of points of ${\rm PG}(7,2)$ and ${\rm PG}(15,2)$ obtained by Bremner and Stavrou~\cite{bs}
 to partition the sets of maximal sets of mutually commuting $N$-qubits operators (for $N=2,3$ and $4$) and provide information on sizes, representatives in ${\rm PG}(2^N-1,2)$, corresponding observables as well as ranks of these sets.

\subsection{Two distinguished classes of mutually commuting two-qubit operators}

It is well known (see, e.g., \cite{hos1}) that the projective space ${\rm PG}(3,2)$ is the union of two $G$-orbits,
$
{\rm PG}(3,2)=\mathcal{O}_1\cup \mathcal{O}_2$,
with $\sharp \mathcal{O}_1=9$ and $\sharp \mathcal{O}_2=6$. The orbit $\mathcal{O}_1$, comprising the points lying on a hyperbolic quadric $\mathcal{Q}^+(3,2)$,
corresponds to the $G$-orbit of any
separable vector in the tensorial basis (for example, the orbit of $x_1^1\otimes x_2^1$). On the other hand, $\mathcal{O}_2$, consisting of six of\/f-quadric points, is the orbit corresponding to non-separable tensors (for example, the orbit of $x_1^0\otimes x_2^0+x_1^1\otimes x_2^1$).
Our bijection associates the two orbits of ${\rm PG}(3,2)$ with two distinguished classes of maximal sets of mutually commuting two-qubit operators, as described in Table~\ref{Table1}.

\begin{table}[!h]\small
 \centering
 \caption{Classes of mutually commuting $2$-qubits operators; here, ${\rm PG}(1,2)_a=\langle XI,IX \rangle$ and ${\rm PG}(1,2)_b=\langle ZX,XZ \rangle$.}\label{Table1}
 \vspace{1mm}
\begin{tabular}{|c|c|c|c|c|c|}
\hline
 Orbit & Size & Representative & Observable & {$T/E$-rank} & $\begin{array}{c} \text{Set of mutually commuting} \\
  \text{two-qubit observables}\end{array}$\\
 \hline
 $\mathcal{O}_1$ & $9$ & $[0:0:1:0]$ & $XI$ & {$1/0$} &${\rm PG}(1,2)_a$\tsep{1pt}\\
 $\mathcal{O}_2$ & $6$ & $[1:0:1:0]$ & $YI$ &{$2/1$} & ${\rm PG}(1,2)_b$\\
 \hline
\end{tabular}
\end{table}

The projective line ${\rm PG}(1,2)_a$, spanned by $\langle XI,IX \rangle$, is obviously mapped to $[0:0:1:0]$ by our construction. Indeed, according to equations~(\ref{corr1}) and (\ref{corr2}), the observables $XI$ and $IX$ correspond to the points of $\mathcal{W}(3,2)$ having the coordinates $(0,0,1,0)$ and $(0,0,0,1)$ and these two points def\/ine the line represented by the matrix
\[
P_a=\begin{pmatrix}
 0 & 0 & 1 & 0\\
 0 & 0 & 0 & 1
\end{pmatrix}\]
whose Pl\"ucker coordinates are $[0:0:1:0]$, i.e., $p_{12}=p_{14}=p_{23}=0$ and $p_{34}=1$.
Similarly, the line def\/ined by
\[P_b=\begin{pmatrix}
 1 & 0 & 0 & 1\\
 0 & 1 & 1 & 0
\end{pmatrix}\]
satisf\/ies $p_{12}=p_{34}=1$ and $p_{14}=p_{23}=0$, i.e., it is mapped to $[1:0:1:0]$. This line is spanned by the points $(1,0,0,1)$ and $(0,1,1,0)$,  being thus generated by $ZX$ and $XZ$.

The partition of ${\rm PG}(3,2)$ into two orbits $\mathcal{O}_1$ and $\mathcal{O}_2$ tells us that we can similarly partition ${\rm LGr}(2,4)$
into two classes of lines; a class of cardinality
$9$,  which is  the $G$-orbit of ${\rm PG}(1,2)_a$, and a class of  cardinality $6$,  which is the $G$-orbit of ${\rm PG}(1,2)_b$.

\subsection{Three distinguished classes of mutually commuting three-qubit operators}

The projective space ${\rm PG}(7,2)$ is the union of f\/ive $G$-orbits (see \cite{bs,hos1,LaS}),
$
{\rm PG}(7,2)=\mathcal{O}_1\cup\mathcal{O}_2\cup\mathcal{O}_3\cup\mathcal{O}_4\cup\mathcal{O}_5$,
with $\sharp\mathcal{O}_1=27$, $\sharp\mathcal{O}_2=54$, $\sharp\mathcal{O}_3=108$, $\sharp\mathcal{O}_4=54$ and $\sharp\mathcal{O}_5=12$.
It is also well known (see, e.g.,~\cite{hos1}) that
$
\mathcal{Q}^+(7,2)=\mathcal{O}_1\cup\mathcal{O}_2\cup\mathcal{O}_4$.
Hence, in light of our main result of Section~\ref{examples3},  the variety $\underline{\pi}({\rm LGr}(3,6))$ is partitioned into three dif\/ferent $G$-orbits whose properties are summarized in Table~\ref{Table2}; here, we used the explicit expression of $\underline{\pi}$ given in~\cite{PMM} and the representatives of the orbits $\mathcal{O}_i$ were taken from~\cite{bs} (transformed, of course, into our adopted system of coordi\-nates).

\begin{table}[!h]\small
 \centering
\caption{Classes of mutually commuting $3$-qubits operators; here, ${\rm PG}(2,2)_a=\langle XII,IXI,IIX \rangle$, ${\rm PG}(2,2)_b=\langle ZZI,XXI,IIX \rangle$, and ${\rm PG}(2,2)_c=\langle XIX,IXX,ZZZ \rangle$.}\label{Table2}
\vspace{1mm}

\begin{tabular}{|c|c|c|c|c|c|}
\hline
 Orbit & Size & Representative & $\begin{array}{@{}c@{}} \text{Symmetric} \\ \text{4-qubit observable}\end{array}$ & {$T/E$-rank} & $\begin{array}{@{}c@{}} \text{Set of mut. commuting} \\
\text{3-qubit observables}\end{array}$\\
 \hline
 $\mathcal{O}_1$ & $27$ & $[0:0:0:0:1:0:0:0]$ & $XIII$ &{$1/0$} & ${\rm PG}(2,2)_a$\tsep{1pt}\\
 $\mathcal{O}_2$ & $54$ & $[0:0:0:1:0:0:1:0]$ & $IIXZ$ &{$2/1$} & ${\rm PG}(2,2)_b$\\
 $\mathcal{O}_4$ & $54$ & $[0:0:0:1:0:1:1:0]$ & $IXXZ$ &{$3/1$} & ${\rm PG}(2,2)_c$\\
 \hline
\end{tabular}
\end{table}

To illustrate how we assign a projective plane of order two to a representative of $\mathcal{O}_i$, let us detail the calculation for the
orbit $\mathcal{O}_2$. A~representative of the second non-trivial orbit in the classif\/ication of \cite{bs} is,
in the tensorial basis, $x_1^0\otimes x_2^1\otimes x_3^1+x_1^1\otimes x_2^0\otimes x_3^1$, which in our notation corresponds to $x_4=x_7=1$. Using the labeling
of the Pl\"ucker coordinates given in Section~\ref{examples3} this means that $p_{156}= p_{246}=1$, the remaining coordinates being zero.
The $3 \times 6$ matrix satisfying these conditions is of the form
\[\left(\begin{array}{@{}ccc|ccc@{}}
                        1 & 1 & 0 & 0 & 0 & 0 \\
                        0 & 0 & 0 & 1 & 1 & 0\\
                        0 & 0 & 0 & 0 & 0 & 1
                              \end{array}\right).\]
The 3 vectors represented by the rows of the matrix are the coordinates of the three points that span
the  ${\rm PG}(2,2)_b$, which is obviously mapped to $[0:0:0:1:0:0:1:0]$.
A~similar partition of ${\rm LGr}(3,6)$ into three dif\/ferent classes is also obvious.

\subsection{Six distinguished classes of mutually commuting four-qubit operators}

The stratif\/ication of ${\rm PG}(15,2)$ in terms of $29$ $G$-orbits was also established in \cite{bs}.
In order to identify the orbits which partition
the variety $\underline{\pi}({\rm LGr}(4,8))$, we checked the representative of each orbit, taken from Table 5 of \cite{bs}, and found out that six of them annihilate the polynomials of the ideal $I$ (see equation~(\ref{ideal})).
The results of our calculations are portrayed in Table~\ref{Table3}  (here the f\/irst non-trivial orbits is denoted $\mathcal{O}_2$ to be compatible with the numbering of~\cite{bs},
which also takes into account the trivial orbit).
\begin{table}[!h]\small\centering
 \caption{Classes of mutually commuting $4$-qubits operators; here,
${\rm PG}(3,2)_a=\langle XIII,IXII,IIXI$, $IIIX \rangle$, ${\rm PG}(3,2)_b=\langle XIII,IXII,IIZZ,IIYY \rangle$, ${\rm PG}(3,2)_c=\langle XIII,IZZZ,IYYZ,IYZY \rangle$, ${\rm PG}(3,2)_d$ $=\langle ZYYY, YZYY,YYZY,YYYZ \rangle$, ${\rm PG}(3,2)_e=\langle XXII,ZZII,IIZZ,IIYY \rangle$ and
${\rm PG}(3,2)_f= \langle XIZZ$, $IXZZ,ZZXI,ZZIX \rangle$.}\label{Table3}
\vspace{1mm}

  \begin{tabular}{|@{\,}c@{\,}|@{\,\,}c@{\,\,}|@{\,\,}c@{\,\,}|@{\,\,}c@{\,\,}|@{\,\,}c@{\,\,}|@{\,}c@{\,}|}
\hline
 Orbit & Size & Representative & $\begin{array}{@{}c@{}} \text{Symmetric} \\ \text{8-qubit obs'le}\end{array}$       &
 $\begin{array}{@{}c@{}} \text{$T/E$-}\\ \text{rank}\end{array}$         & $\begin{array}{@{}c@{}} \text{Set of mut. comm.} \\ \text{4-qubit obs'les}\end{array}$\\
 \hline
 $\mathcal{O}_2$ & $81$ &     $[0:0:0:0:0:0:0:0:1:0:0:0:0:0:0:0]$ & $XIIIIIII$&{$1/0$}  & ${\rm PG}(3,2)_a$\tsep{1pt}\\
 $\mathcal{O}_3$ & $324$ &    $[0:0:0:0:0:0:0:0:0:1:1:0:0:0:0:0]$ & $IXXIIIII$&{$2/1$} & ${\rm PG}(3,2)_b$\\
 $\mathcal{O}_6$ & $648$ &    $[0:0:0:0:0:0:0:0:0:1:1:0:1:0:0:0]$ & $IXXIXIII$&{$3/1$} & ${\rm PG}(3,2)_c$\\
 $\mathcal{O}_{14}$ & $162$ & $[0:0:0:0:0:0:0:1:0:1:1:0:1:0:0:0]$ & $IXXIXIIZ$&{$4/1$} & ${\rm PG}(3,2)_d$\\
 $\mathcal{O}_{17}$ & $108$ & $[0:0:0:0:0:1:1:0:0:0:0:0:0:1:1:0]$ & $IIIIIYYI$&{$4/2$} & ${\rm PG}(3,2)_e$\\
 $\mathcal{O}_{18}$& $972$ &  $[0:0:0:0:0:1:1:0:1:0:0:0:0:1:1:0]$ & $XIIIIYYI$&{$4/2$} & ${\rm PG}(3,2)_f$\\
 \hline
 \end{tabular}
 \end{table}

 To identify the ${\rm PG}(3,2)$'s that correspond to the representatives of the orbits we proceed similarly as in the previous two cases, that is, we create the $4\times 8$ matrix whose minors satisfy
 the conditions implied by the corresponding representative.
 ${\rm LGr}(4,8)$ is likewise  partitioned into six non-equivalent classes.

\section{The Lagrangian Grassmannian and the variety\\ of principal minors}\label{variety_minor}

At this point is is worth mentioning several papers \cite{LS, O1,O2} that deal with similar problems over the f\/ield of complex numbers and which are deeply related to the  construction over ${\rm GF}(2)$ considered in this paper.

Let $\KK=\CC$ and let $\mathcal{Z}_N\subset \PP(\underbrace{\CC^2\otimes\dots\otimes\CC^2}_{N \text{ times}})$ be the image of the following
rational map \cite{O2}:
 \begin{alignat*}{5}
&\phi: \  && \PP\big(S^2 \CC^n\oplus \CC\big) && \dashrightarrow \ && \PP\big(\big(\CC^2\big)^{\otimes N}\big),&\\
      &&& [A,t] && \mapsto && \big[t^{n-|I|} \Delta_I(A)X^I\big], &
\end{alignat*}
with $A$ being a symmetric complex matrix and $X^I=x_1 ^{i_1}\otimes\cdots \otimes x_N ^{i_N}$, $i_j= \begin{cases}
                                                                                                           0 & \text{if}\ j\notin I,\\
                                                                                                           1 & \text{ if} \ j\in I,
                                                                                                          \end{cases}$  a~tensorial basis of $(\CC^2)^{\otimes N}$.
$\mathcal{Z}_N$ is an algebraic variety, called the variety of principal minors of symmetric matrices, corresponding to the linear projection of the Lagrangian Grassmannian (over the complex numbers).
The linear projection means that from the set of minors of cardinality $\binom{2N}{N}$ we only keep the set of principal minors of cardinality $2^N$, i.e.\ it is the type of
the projection $\underline{\pi}$ def\/ined in Section~\ref{bijection} over ${\rm GF}(2)$. However, in the complex case this projection is no
longer a bijection, that is, the principal minors do not contain all the information on the
Lagrangian Grassmannian.  In particular, over $\CC$, as well as over any algebraically closed f\/ield of characteristic dif\/ferent from two,
the of\/f-diagonal entries of the symmetric matrices (Section~\ref{bijection}) are determined by the principal minors ($a_{ij}^2=\Delta_i(A)\Delta_j(A)-\Delta_{i,j}(A)$) only up to
the sign and, thus, the projection is generically two to one.

\looseness=-1
The motivation for studying $\mathcal{Z}_N$ in the complex case comes from the principal minors assignment problem \cite{HS,HSt,LS}. This
problem asks for  necessary and suf\/f\/icient conditions
for a collection of $2^N$ numbers to arise as the principal minors of an $N\times N$ matrix.
In the case of a symmetric matrix, a collection of $2^N$ numbers corresponds to its principal minors if and only of the
corresponding point in~$\PP((\CC^2)^{\otimes N})$ belongs to the variety $\mathcal{Z}_N$.
This problem for symmetric matrices was solved by Oeding~\cite{O1,O2}, who successfully described a
set of degree-four polynomials which cut out the variety~$\mathcal{Z}_N$.
In particular, Oeding proved, using representation theory techniques, that a set of equations def\/ining $\mathcal{Z}_N$ are obtained by taking the
$G=SL_2(\CC)\times\dots\times SL_2(\CC)\ltimes S_N$ orbit of a certain peculiar quartic polynomial (the so-called $2\times 2\times 2$
Cayley hyperdeterminant). Oeding's result also provides a set-theoretical solution to a conjecture of Holtz and Strumfels~\cite{HSt} which says that the $G$-orbit of the Cayley hyperdeterminant generates the ideal of the variety~$\mathcal{Z}_N$.

It is, naturally, tempting to rephrase Oeding's result {and the conjecture of Holtz and Strumfels} into the ${\rm GF}(2)$-regime and check if one can recover the equations obtained in Sections~\ref{examples3} and~\ref{examples4}.
Over ${\rm GF}(2)$, the equations provided by Corollary~1.4 of~\cite{O2} lead to the ${\rm SL}(2,2)\times \cdots\times {\rm SL}(2,2)\ltimes S_N$ orbit of
\[
\mathcal{Q}=p_{{\bf 123}4\dots N}p_{{\bf\overline{1} \overline{2}\overline{3}}4\dots N}+p_{{\bf 12\overline{3}}4\dots N}p_{{\bf 3 \overline{1}\overline{2}}4\dots N}+p_{{\bf 13\overline{2}}4\dots N}p_{{\bf 2\overline{1}\overline{3}}4\dots N}+p_{{\bf 1\overline{2}\overline{3}}4\dots N}p_{{\bf 23\overline{1}}4\dots N},
\]
where $\overline{k}=N+k$. \looseness=-1
For $N=3$, $\mathcal{Q}=0$ is readily recognized to be identical to equation~(\ref{quadric36}) def\/ining $\underline{\pi}({\rm LGr}(3,6))$.
 The quadratic polynomial $\mathcal{Q}$ of equation~(\ref{quadric36})  is nothing but the irreducible factor of
the Cayley hyperdeterminant which can be written as $\mathcal{Q}^2$ over ${\rm GF}(2)$ (see Remark~18 of~\cite{K}).  Thus,
for $N=3$, both Oeding's result and
Holtz and Strumfels' conjecture are true over~${\rm GF}(2)$.
For $N=4$, the distinguished polynomial $\mathcal{Q}$  coincides with our $\mathcal{Q}_8$ appearing  in the ideal def\/ined by equation~(\ref{ideal}).
It can be shown that the $G$-orbit of $\mathcal{Q}=\mathcal{Q}_8$  consists of the polynomials
$\mathcal{Q}_1, \mathcal{Q}_2,\dots,\mathcal{Q}_8$ and  $\mathcal{Q}_0$. However,  via the repeated action of the generators of
$G$  we did not manage to get the remaining four-term polynomials $\mathcal{Q}_9$ and $\mathcal{Q}_{10}$, merely their
sum $\mathcal{Q}_0=\mathcal{Q}_9+\mathcal{Q}_{10}$. { This means that in the $N=4$ case, the  $G$-orbit of the Cayley hyperdeterminant over ${\rm GF}(2)$ does not generate the ideal
of the variety of principal minors,  i.e.\ the Holtz--Strumfels conjecture is not valid. However, Oeding's result remains
true, for one can readily check that the set $\mathcal{Q}_1, \mathcal{Q}_2,\dots, \mathcal{Q}_9$ and $\mathcal{Q}_0$ indeed cuts out the variety} $\underline{\pi}({\rm LGr}(4,8))$.
This case thus features over ${\rm GF}(2)$ some subtle properties that have no counterpart over the f\/ield of complex numbers.
Nevertheless, we are convinced that the approach developed by Oeding is a very promising one, which can be appropriately adjusted/modif\/ied to be meaningful also over the smallest Galois f\/ield.

\section{Conclusion}

In this paper, we gave, for any $N \geq 2$, a rigorous proof of the existence of a bijection between the set of generators of the symplectic polar space $\mathcal{W}(2N-1,2)$
and a distinguished subset of points of $\mathcal{W}(2^N-1,2)$. Physically speaking, we established a one-to-one mapping between the maximal sets of pairwise commuting operators
of the $N$-qubit Pauli group and a  subset of the  $2^{N-1}$ qubit observables.  Proving this correspondence, we also found a method how to get the def\/ining equations of the image of
the mapping within ${\rm PG}(2^N-1,2)$ and explicitly illustrated this method for the cases $N=2$, $N=3$ and $N=4$.

The image of our mapping is over the complex numbers known as the variety of principal minors of symmetric matrices \cite{O1,O2}.  We have also pointed out that
the calculations in the ${\rm GF}(2)$-regime deserve a special treatment and are, in general, not the direct translation of the results obtained over the f\/ield of the complex numbers.
Since ${\rm GF}(2)$-settings have already acquired a f\/irm footing in the context of Quantum Information Theory,
we aim at getting deeper insights into the variety of principal minors of ${\rm GF}(2)$-symmetric matrices employing, if possible, a more coordinate-free approach.

\subsection*{Acknowledgments}
This work began while two of the authors (M.S.~and~P.L.) were fellows of the ``Research in Pairs'' Program of the Mathematisches Forschungsinstitut Oberwolfach (Oberwolfach, Germany), in the period from 24 February to 16 March, 2013.
It was also partially supported by the PEPS ICQ 2013 project CoGIT of the CNRS (F.H. and M.S.), the VEGA Grant Agency, grant No.~2/0003/13 (M.S.) and by the MTA-BME Condensed Matter Physics Research Group, grant No.~04119 (P.L.).

\pdfbookmark[1]{References}{ref}
\LastPageEnding


\begin{thebibliography}{99}
\footnotesize\itemsep=0pt

\bibitem{aft}
Alexeev B., Forbes M.A., Tsimerman J., Tensor rank: some lower and upper
  bounds, in 26th {A}nnual {IEEE} {C}onference on {C}omputational {C}omplexity,
  IEEE Computer Soc., Los Alamitos, CA, 2011, 283--291, \href{http://arxiv.org/abs/1102.0072}{arXiv:1102.0072}.

\bibitem{bs}
Bremner M.R., Stavrou S.G., Canonical forms of {$2\times 2\times 2$} and
  {$2\times 2\times 2\times 2$} arrays over {$\mathbb{F}_2$} and
  {$\mathbb{F}_3$}, \href{http://dx.doi.org/10.1080/03081087.2012.721362}{\textit{Linear Multilinear Algebra}} \textbf{61} (2013),
  986--997, \href{http://arxiv.org/abs/1112.0298}{arXiv:1112.0298}.

\bibitem{cam}
Cameron P.J., Projective and polar spaces, \textit{QMW Mathematics Notes},
  Vol.~13, Queen Mary and Westf\/ield College, London, 1991, available at
  \url{http://www.maths.qmul.ac.uk/~pjc/pps/}.

\bibitem{CZ}
Carrillo-Pacheco J., Zaldivar F., On {L}agrangian--{G}rassmannian codes,
  \href{http://dx.doi.org/10.1007/s10623-010-9434-4}{\textit{Des. Codes Cryptogr.}} \textbf{60} (2011), 291--298.

\bibitem{CLO}
Cox D., Little J., O'Shea D., Ideals, varieties, and algorithms. An
  introduction to computational algebraic geometry and commutative algebra, 3rd~ed., \href{http://dx.doi.org/10.1007/978-0-387-35651-8}{\textit{Undergraduate Texts in Mathematics}}, Springer, New York, 2007.

\bibitem{GKZ}
Gel'fand I.M., Kapranov M.M., Zelevinsky A.V., Discriminants, resultants, and
  multidimensional determinants, \href{http://dx.doi.org/10.1007/978-0-8176-4771-1}{\textit{Mathematics: Theory \& Applications}},
  Birkh\"auser Boston, Inc., Boston, MA, 1994.

\bibitem{hos}
Havlicek H., Odehnal B., Saniga M., Factor-group-generated polar spaces and
  (multi-)qudits, \href{http://dx.doi.org/10.3842/SIGMA.2009.096}{\textit{SIGMA}} \textbf{5} (2009), 096, 15~pages,
  \href{http://arxiv.org/abs/0903.5418}{arXiv:0903.5418}.

\bibitem{hos1}
Havlicek H., Odehnal B., Saniga M., On invariant notions of {S}egre varieties
  in binary projective spaces, \href{http://dx.doi.org/10.1007/s10623-011-9525-x}{\textit{Des. Codes Cryptogr.}} \textbf{62}
  (2012), 343--356, \href{http://arxiv.org/abs/1006.4492}{arXiv:1006.4492}.

\bibitem{HS}
Holtz O., Schneider H., Open problems on {GKK} {$\tau$}-matrices,
  \href{http://dx.doi.org/10.1016/S0024-3795(01)00492-X}{\textit{Linear Algebra Appl.}} \textbf{345} (2002), 263--267,
  \href{http://arxiv.org/abs/math.RA/0109030}{math.RA/0109030}.

\bibitem{HSt}
Holtz O., Sturmfels B., Hyperdeterminantal relations among symmetric principal
  minors, \href{http://dx.doi.org/10.1016/j.jalgebra.2007.01.039}{\textit{J.~Algebra}} \textbf{316} (2007), 634--648,
  \href{http://arxiv.org/abs/math.RA/0604374}{math.RA/0604374}.

\bibitem{K}
Krutelevich S., Jordan algebras, exceptional groups, and {B}hargava
  composition, \href{http://dx.doi.org/10.1016/j.jalgebra.2007.02.060}{\textit{J.~Algebra}} \textbf{314} (2007), 924--977,
  \href{http://arxiv.org/abs/math.NT/0411104}{math.NT/0411104}.

\bibitem{L}
Landsberg J.M., Tensors: geometry and applications, \textit{Graduate Studies in
  Mathematics}, Vol.~128, Amer. Math. Soc., Providence, RI, 2012.

\bibitem{LaS}
Lavrauw M., Sheekey J., Orbits of the stabiliser group of the {S}egre variety
  product of three projective lines, \href{http://dx.doi.org/10.1016/j.ffa.2013.11.002}{\textit{Finite Fields Appl.}} \textbf{26}
  (2014), 1--6, \href{http://arxiv.org/abs/1207.3972}{arXiv:1207.3972}.

\bibitem{PMM}
L\'evay P., Planat M., Saniga M., Grassmannian connection between three-and
  four-qubit observables, Mermin's contextuality and black holes,
  \href{http://dx.doi.org/10.1007/JHEP09(2013)037}{\textit{J.~High Energy Phys.}} \textbf{2013} (2013), no.~9, 037, 35~pages,
  \href{http://arxiv.org/abs/1305.5689}{arXiv:1305.5689}.

\bibitem{LS}
Lin S., Sturmfels B., Polynomial relations among principal minors of a
  {$4\times 4$}-matrix, \href{http://dx.doi.org/10.1016/j.jalgebra.2009.06.026}{\textit{J.~Algebra}} \textbf{322} (2009), 4121--4131,
  \href{http://arxiv.org/abs/0812.0601}{arXiv:0812.0601}.

\bibitem{mer}
Mermin N.D., Hidden variables and the two theorems of {J}ohn {B}ell,
  \href{http://dx.doi.org/10.1103/RevModPhys.65.803}{\textit{Rev. Modern Phys.}} \textbf{65} (1993), 803--815.

\bibitem{O1}
Oeding L., G-varieties and the principal minors of symmetric matrices, Ph.D.\
  Thesis, Texas A\&M University, 2009.

\bibitem{O3}
Oeding L., Set-theoretic def\/ining equations of the tangential variety of the
  {S}egre variety, \href{http://dx.doi.org/10.1016/j.jpaa.2010.09.009}{\textit{J.~Pure Appl. Algebra}} \textbf{215} (2011),
  1516--1527, \href{http://arxiv.org/abs/0911.5276}{arXiv:0911.5276}.

\bibitem{O2}
Oeding L., Set-theoretic def\/ining equations of the variety of principal minors
  of symmetric matrices, \href{http://dx.doi.org/10.2140/ant.2011.5.75}{\textit{Algebra Number Theory}} \textbf{5} (2011),
  75--109, \href{http://arxiv.org/abs/0809.4236}{arXiv:0809.4236}.

\bibitem{pla}
Planat M., Pauli graphs when the {H}ilbert space dimension contains a square:
  why the {D}edekind psi function?, \href{http://dx.doi.org/10.1088/1751-8113/44/4/045301}{\textit{J.~Phys.~A: Math. Theor.}}
  \textbf{44} (2011), 045301, 16~pages, \href{http://arxiv.org/abs/1009.3858}{arXiv:1009.3858}.

\bibitem{ps}
Planat M., Saniga M., On the {P}auli graphs on {$N$}-qudits, \textit{Quantum
  Inf. Comput.} \textbf{8} (2008), 127--146, \mbox{\href{http://arxiv.org/abs/quant-ph/0701211}{quant-ph/0701211}}.

\bibitem{slp}
Saniga M., L{\'e}vay P., Pracna P., Charting the real four-qubit {P}auli group
  via ovoids of a hyperbolic quadric of {${\rm PG}(7,2)$}, \href{http://dx.doi.org/10.1088/1751-8113/45/29/295304}{\textit{J.~Phys.~A:
  Math. Theor.}} \textbf{45} (2012), 295304, 16~pages, \href{http://arxiv.org/abs/1202.2973}{arXiv:1202.2973}.

\bibitem{sp}
Saniga M., Planat M., Multiple qubits as symplectic polar spaces of order two,
  \textit{Adv. Stud. Theor. Phys.} \textbf{1} (2007), 1--4,
  \href{http://arxiv.org/abs/quant-ph/0612179}{quant-ph/0612179}.

\bibitem{spp}
Saniga M., Planat M., Prachna P., Projective curves over a ring that includes
  two-qubits, \href{http://dx.doi.org/10.1007/s11232-008-0076-x}{\textit{Theoret. and Math. Phys.}} \textbf{155} (2008), 905--913,
  \href{http://arxiv.org/abs/quant-ph/0611063}{quant-ph/0611063}.


\bibitem{thas}
Thas J.A., Ovoids and spreads of f\/inite classical polar spaces, \href{http://dx.doi.org/10.1007/BF01447417}{\textit{Geom.
  Dedicata}} \textbf{10} (1981), 135--143.

\bibitem{th}
Thas K., The geometry of generalized Pauli operators of $N$-qudit Hilbert
  space, and an application to MUBs, \href{http://dx.doi.org/10.1209/0295-5075/86/60005}{\textit{Europhys. Lett.}} \textbf{86}
  (2009), 60005, 3~pages.

\bibitem{wae}
Waegell M., Primitive nonclassical structures of the $N$-qubit Pauli group,
  \href{http://arxiv.org/abs/1310.3419}{arXiv:1310.3419}.

\end{thebibliography}
\end{document}